\documentclass[aps,showpacs,prl,amssymb, amsmath, twocolumn]{revtex4-1}
\usepackage{graphicx}
\usepackage{amssymb}
\usepackage{amsmath}
\usepackage{bm}
\usepackage{xcolor} 
\usepackage{mathtools}

\newcommand{\bra}[1]{\left\langle #1\right|}
\newcommand{\ket}[1]{\left|#1\right\rangle}

\begin{document}
\title{Rydberg atom entanglements in the weak coupling regime}
\author{Hanlae Jo, Yunheung Song, Minhyuk Kim, and Jaewook Ahn}
\email{jwahn@kaist.ac.kr}
\address{Department of Physics, KAIST, Daejeon 305-701, Korea}

\begin{abstract}
We present an entanglement scheme for Rydberg atoms using the van der Waals interaction phase induced by Ramsey-type pulsed interactions. This scheme realizes not only controlled phase operations between atoms at a distance larger than Rydberg blockade distance, but also various counter-intuitive entanglement examples, including two-atom entanglement in the presence of a closer third atom and $W$-state generation for partially-blockaded three atoms. Experimental realization is conducted with single rubidium atoms loaded in an array of optical tweezer dipole traps, to demonstrate the proposed entanglement generations and measurements.
 \end{abstract}

\maketitle
Quantum entanglement is one of the most bizarre and intriguing natures of quantum mechanics~\cite{Horodecki2009}, which plays an important role in understanding the physics of quantum many-body systems~\cite{Islam2015, Ryd_simulator_lukin, Ryd_simulator_HSK} and also empowering various quantum applications such as quantum computing~\cite{Nielsen2010}, quantum sensing~\cite{Degen2017}, and quantum communications~\cite{Ursin2007}. Currently, there is a strong interest for the generation, manipulation, and detection of quantum entanglements, demonstrated in many physical systems including photons~\cite{entangled_photon}, 
atoms~\cite{wilk2010, Isenhower2010, jau2016, high_fidelity}, ions~\cite{entanglement_ion}, and solid-state systems such as superconducting circuits~\cite{cnot_supercond} and defective diamonds~\cite{NV_entanglement}. However, in these systems, entanglement skills need much improvement yet, even to operate a small-scale quantum computer. 

Entanglement of arbitrary qubit pairs, especially ones that are not in the proximity, is of particular importance for a scalable quantum system of good connectivity. Although it has been achieved, for example, in trapped ions by common-mode motion~\cite{Cirac_zoller1, Cirac_zoller2} and in superconducting circuits by cavity bus~\cite{entalgment_supercond}, it has not been realized in most other systems including Rydberg-atom systems, of particular relevance in the context of the present paper. The widely-used entanglement scheme of Rydberg atom systems~\cite{wilk2010, Isenhower2010,  jau2016, high_fidelity} is based on Rydberg-blockade effect~\cite{Jaksch2000}, which prohibits double excitation to a Rydberg energy state among atoms closer than the blockade radius $r_b = (C_6/\Omega)^{1/6}$ defined by Rabi frequency $\Omega$ and van der Waals interaction strength $C_6$. In this scheme, however, all the pairs of atoms within the blockade radius are to be simultaneously entangled, making selective entanglement, of an arbitrary pair of atoms, difficult.

In this paper, we experimentally demonstrate  atom-pair entanglement in the weak-coupling regime, which is closely related to the model A in Ref.~\cite{Jaksch2000}. With this, pair-wise atom entanglements are enabled beyond the Rydberg blockade distance, even in the presence of closer atoms that are to be left unentangled. In the weak-coupling regime ($d>r_b$), the doubly-excited Rydberg state of two atoms separated by a distance $d$ gains interaction phase $\alpha=\tau {C_6}/{d^6}$, i.e., $|11\rangle \rightarrow \exp({-i \alpha})|11\rangle$, during an interaction time $\tau$, where the pseudo-spin states $|1\rangle$  and $|0\rangle$ represent the Rydberg and ground states, respectively. So, a pair of atoms separated by $d_\pi=(\tau {C_6}/\pi)^{1/6}$ undergoes a controlled $\pi$-phase gate and $d_{2\pi}=(\tau {C_6}/2\pi)^{1/6}$ a controlled $2\pi$-phase gate that is the Null gate. Utilizing this interaction phase, we present three entanglement examples: First, we generate the entanglement of two atoms at a distance beyond the blockade radius. Second, we use three atoms in the linear configuration, $ABC$, with $d_{AB}>d_{BC}>r_b$, and operate the controlled $\pi$-phase gate only on the $AB$ pair, while the closer pair $BC$ are left separable. Third, we produce the $W$-state of partially blockaded three atoms using a two-pulse coherent control scheme.

\begin{figure}[tbp]
\centering
 \includegraphics[width=0.45\textwidth]{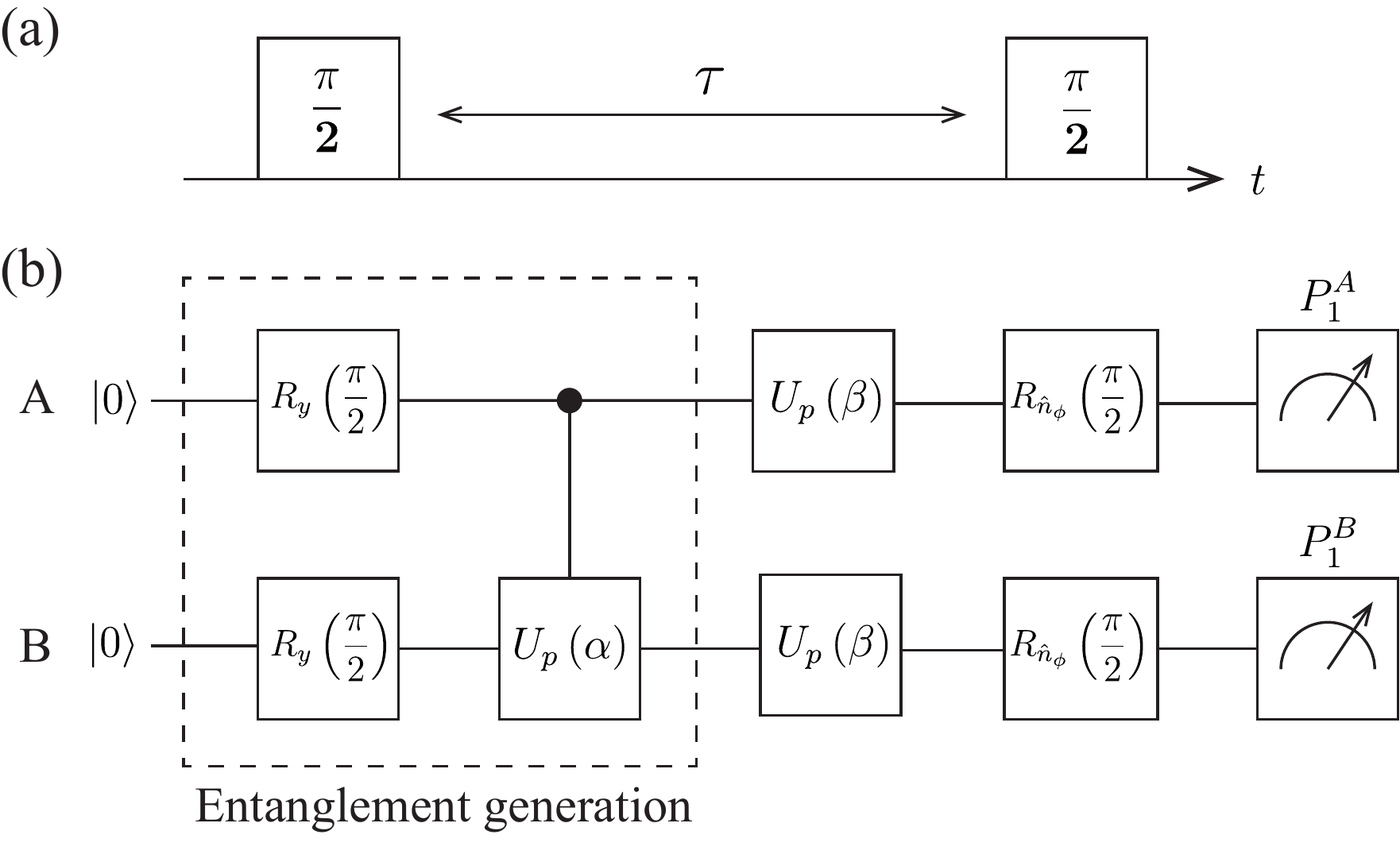}
 \caption{(a) Ramsey-type double $\pi/2$-interactions. (b) Corresponding quantum circuit, in which two atoms $A$ and $B$ undergo the first Ramsey interaction $R_{\hat{y}}({\pi}/{2})$, controlled $U_p(\alpha=\tau C_6/d^6)$ phase gate, a residual local phase $U_p(\beta)$, and the second Ramsey $R_{\hat{n}_\phi}({\pi}/{2})$ fine-controlled by laser phase $\phi$.}
\label{fig1}
\end{figure}

We use Ramsey interferometry to generate and measure the two-atom entanglements in the weak coupling regime. As in Fig.~\ref{fig1}(a), two resonant $\pi/2$ pulses time-separated by $\tau$ interact two atoms $A$ and $B$ separated at a distance $d$ ($> r_b$) and initially in the ground state $|00\rangle$. The first Ramsey pulse rotates each atom about the $y$-axis to prepare the superposition state $\ket{++}$, in which $\ket{+}=(\ket{0}+\ket{1})/\sqrt{2}$ of each atom. During the time interval $\tau$, van der Waals interaction induces a phase factor $e^{-i\alpha}$ to the doubly-excited state $\ket{11}$. Finally the second Ramsey pulse rotates each atom by $\pi/2$ about the axis $\hat{n}_\phi=\cos\phi\hat{y}-\sin\phi\hat{x}$, where $\phi$ is the laser phase that fine-controls the azimuthal angle of the Bloch vector with respect to the $y$-axis. The given optical process is described by the following unitary operation defined for the two-qubit Hilbert space as: 
\begin{eqnarray}
U(\alpha,\phi) = R_{\hat{n}_\phi,A}^{\pi/2}\otimes R_{\hat{n}_\phi,B}^{\pi/2} e^{-i  n_A n_B \alpha } R_{\hat{y},A}^{\pi/2} \otimes R_{\hat{y},B}^{\pi/2},
\label{eq1}
\end{eqnarray}
where $R_{\hat{y} }^{\pi/2}$ and $R_{\hat{n}_\phi }^{\pi/2}$ are single-qubit $\pi/2$ rotations about the axes, $\hat{y}$ and $\hat{n}_\phi$, respectively; and $n_{A,B}$ are the excitation number of the atoms. An equivalent quantum circuit is shown in Fig.~\ref{fig1}(b), including an additional state-independent local phase $\beta$ which will be explained in the experiment. After the entanglement generation enclosed by a dashed box, the two-atom state evolves to
\begin{equation}
|\psi(\tau)\rangle= \frac{1}{2} \left( \ket{00} 
+ \ket{01} + \ket{10} + e^{-i\alpha}\ket{11} \right),
\label{eq2}
\end{equation}
which is maximally entangled at $\alpha = (2n+1) \pi$, for an integer $n$, or loses entanglement at $\alpha=2n\pi$. The resulting entanglement can be characterized by a Ramsey-type measurement~\cite{ entanglement_measurement2} with the second $\pi/2$-pulse. As a function of the control phase $\phi$, the probability of each atom, projected to $\ket{1}$, is given by
\begin{equation}
P_1^A = P_1^B= \frac{1}{2}+\frac{1}{2}\cos\left(\frac{\alpha}{2}\right)\cos\left(\frac{\alpha}{2}+\beta +\phi  \right).
\label{ramsey}
\end{equation}
So, the resulting fringe visibility, $\cos(\alpha/2)$, manifests the entanglement: maximal (minimal) visibility for no (maximal) entanglement. 

Experiments were performed with an apparatus previously reported elsewhere~\cite{Ryd_simulator_HSK, tweezer, WJL2019}. In brief, we used optical tweezers to trap rubidium ($^{87}$Rb) single atoms and, with a 2-ms optical pumping, prepared them in the ground state $\ket{0}=|{{\rm 5S}_{1/2}, F=2, m_F = 2}\rangle$. We then turned off the optical tweezers and excited the atoms to a Rydberg state $\ket{1}=|{{\rm 67S}_{1/2}, J = {1/2}, m_J = 1/2}\rangle$, through the off-resonant intermediate state $|{\rm 5P}_{3/2}\rangle$, with counter-propagating 780-nm and 480-nm beams of $\sigma^{+}$ and $\sigma^{-}$ polarizations, respectively. With $C_6= 2\pi \times 513$~GHz~$\mu$m$^6$~\cite{pairint}, two-photon Rabi frequency $\Omega = 2\pi \times 0.83$~MHz gives the blockade radius of $r_b=9.23$~$\mu$m in the first two experiments. So, a time delay $\tau=2.6$~$\mu$s renders $d_\pi=1.28r_b$ for $\alpha=\pi$ and $d_{2\pi}=1.14r_b$ for $\alpha=2\pi$. The repulsive force between Rydberg atoms is negligible in the experiments; atom displacement by the repulsive force during a controlled $\pi$-phase gate operation is estimated less than 4~nm, i.e., $\Delta \alpha/\alpha <0.2\%$.

The Ramsey pulses were produced with an acousto-optic modulator by switching the 780-nm beam on and off twice, while the 480-nm beam was left on. The two-photon resonance was maintained with the laser-induced AC Stark shift, $\delta_{AC}=2.1$~MHz, taken into account. So, during the time interval, the excited state of each atom gained the phase $\beta= 2\pi \times \delta_{AC} \tau$, while the doubly-excited $\ket{11}$ state gained $\alpha$, as in Eq.~\eqref{eq2}.
 After the interactions of the two $\pi/2$ pulses, the optical tweezers were turned back on to recapture the ground-state atoms, which were then recorded through fluorescence imaging of the cyclic transition between $|{{\rm 5S}_{1/2}, F=2}\rangle$ and $|{{\rm 5P}_{3/2}, F'=3}\rangle$.  About 200-400 times of measurements were accumulated for $P_1(\phi)$, and the entire experiments were repeated by varying the laser phase $\phi$.
 
 \begin{figure}[tbp]
\centering
 \includegraphics[width=0.48\textwidth]{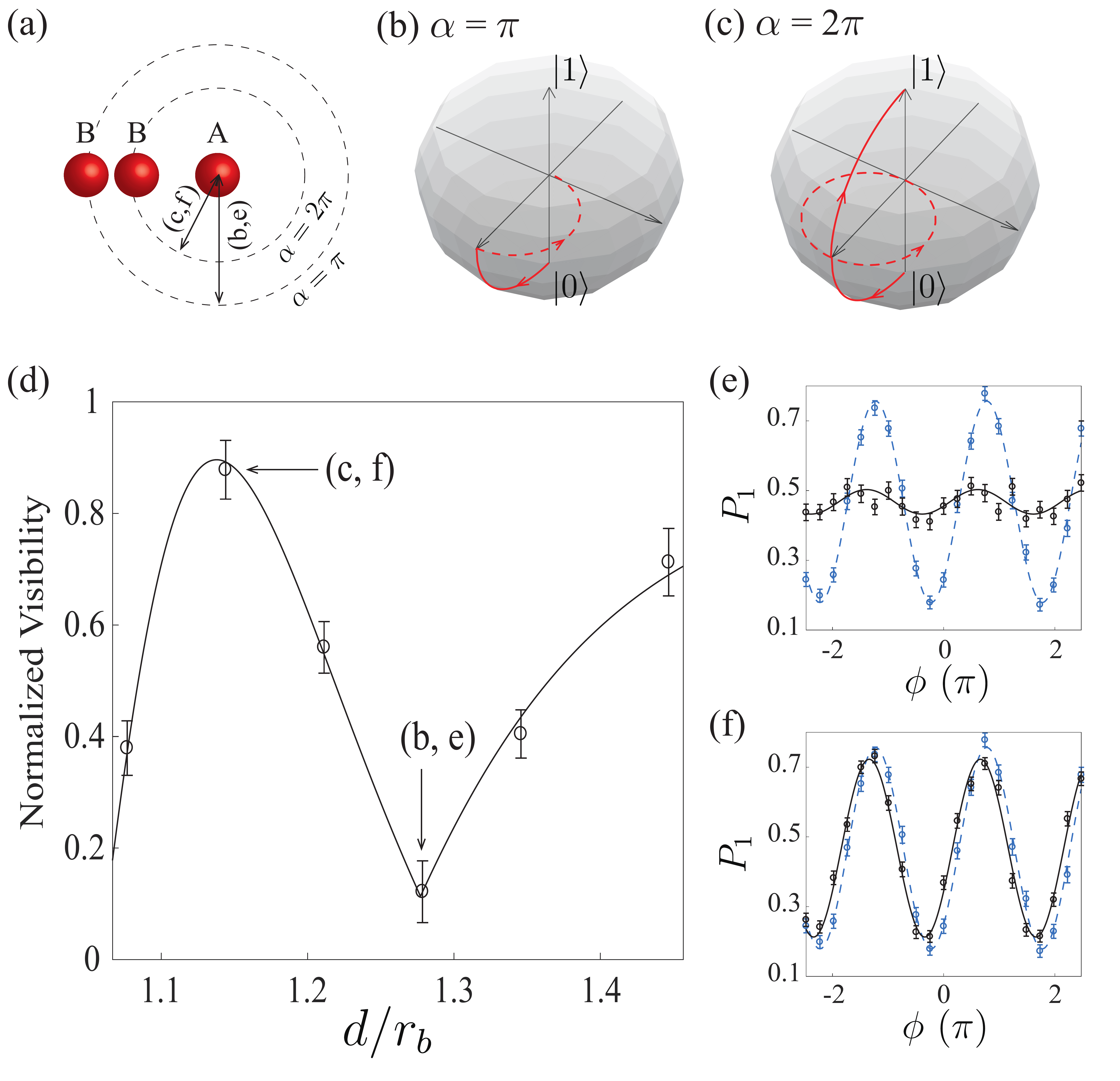}
 \caption{(a) Two atoms $A$ and $B$ separated at a distance of either $d_{\pi}$ for $\alpha=\pi$ or $d_{2\pi}$ for $\alpha=2\pi$.
(b,c) Quantum trajectories of each atom expected for (b) $\alpha = \pi$ and (c) $2\pi$. (d) Ramsey fringe visibilities measured at various distances compared with the theoretical line $|\cos(\tau C_6/2d^6)|$ in Eq.~\eqref{ramsey} (scaled and up-shifted for clarity). 
(e,f) Measured Ramsey fringes, $P_1(\phi)=(P_1^A+P_1^B)/2$, for (e) $\alpha = \pi$ and (f) $\alpha=2\pi$, compared with an isolated single-atom Ramsey fringe in blue.}
 \label{fig2}
\end{figure}

Figure~\ref{fig2} summarizes the result of the first experiment, atom-pair entanglements in the weak coupling regime. Two atoms $A$ and $B$ were placed at a distance of either $d_\pi=1.28r_b$ for maximal entanglement or $d_{2\pi}=1.14r_b$ for no entanglement, as shown in Fig.~\ref{fig2}(a). Expected quantum trajectories, of each atom, are plotted on the Bloch sphere in Figs.~\ref{fig2}(b) and \ref{fig2}(c), respectively. In the maximal entanglement case in Fig.~\ref{fig2}(b), the quantum state of each atom $A$ or $B$ evolves from $\ket{0}$ to $\ket{+}$, due to the first Ramsey interaction, and then to the center of the Bloch sphere, driven by the van der Waals interaction of $\alpha$ = $\pi$, so the second Ramsey makes no change to the atom and, as a result, the Ramsey fringe disappears. While, in the no entanglement case in Fig.~\ref{fig2}(c), the interaction of $\alpha$ = $2\pi$ makes each atom return to $\ket{+}$, so the Ramsey fringe is maximally expected. Measured fringe visibilities are plotted for six different distances in Fig.~\ref{fig2}(d), in which the maximal and minimal visibilities correspond to $\alpha=\pi$ (b) and $2\pi$ (c), respectively. Measured Ramsey fringes for $\alpha= \pi$ and $2\pi$ are respectively shown in Figs.~\ref{fig2}(e) and \ref{fig2}(f), agreeing well with the expectations.

\begin{figure}[tbp]
 \centering
 \includegraphics[width=0.48\textwidth]{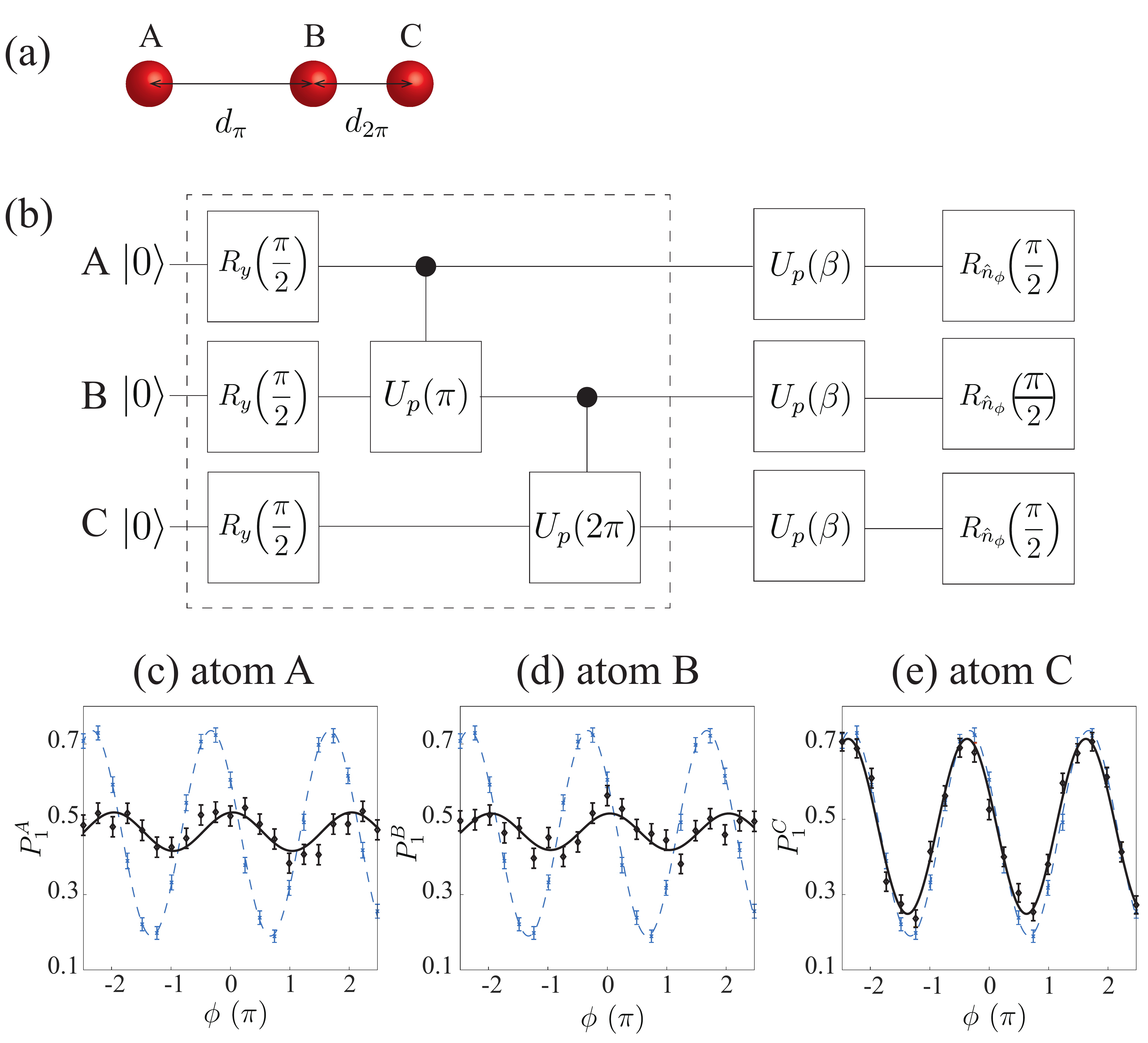}
 \caption{ (a) Linear arrangement of three atoms $ABC$ with $d_{AB}=d_\pi$ and $d_{BC}=d_{2\pi}$. (b) An equivalent three-qubit quantum circuit. (c-e) Measured Ramsey fringes $P_1(\phi)$ of (c)  atom $A$, (d) atom $B$, and (f) atom $C$, in comparison with that of an isolated single-atom in blue.}
 \label{fig3}
\end{figure}

The above scheme of pair-wise entanglements works even in the presence of additional closer atoms, as long as they are properly placed. As an example, we consider a linear configuration of three atoms ($A$, $B$, and $C$), as shown in Fig.~\ref{fig3}(a), in which $A$ and $B$ are placed at $d_{AB}=d_\pi$ and the third atom $C$ satisfies $d_{BC}=d_{2\pi}$. Then, $d_{AC}=d_\pi+d_{2\pi}$ approximates $\alpha_{AC}\approx 0$. The corresponding quantum circuit is drawn in Fig.~\ref{fig3}(b), which includes a controlled $\pi$-phase gate between $A$ and $B$, and a controlled $2\pi$-phase gate between $B$ and $C$. Ramsey measurements are shown in Figs.~\ref{fig3}(c,d,e) for the atoms $A$, $B$, and $C$, respectively. Entangled atoms, $A$ and $B$, exhibit low fringe visibilities of 19$\%$ and 18$\%$, respectively, while the un-entangled atom $C$ a high fringe visibility, closer to the case of an isolated single atom. Therefore, the result indicates that the remote pair ($AB$) can be entangled, while the closer pair ($BC$) is left un-entangled.

\begin{figure}[tbp]
 \centering
 \includegraphics[width=0.48\textwidth]{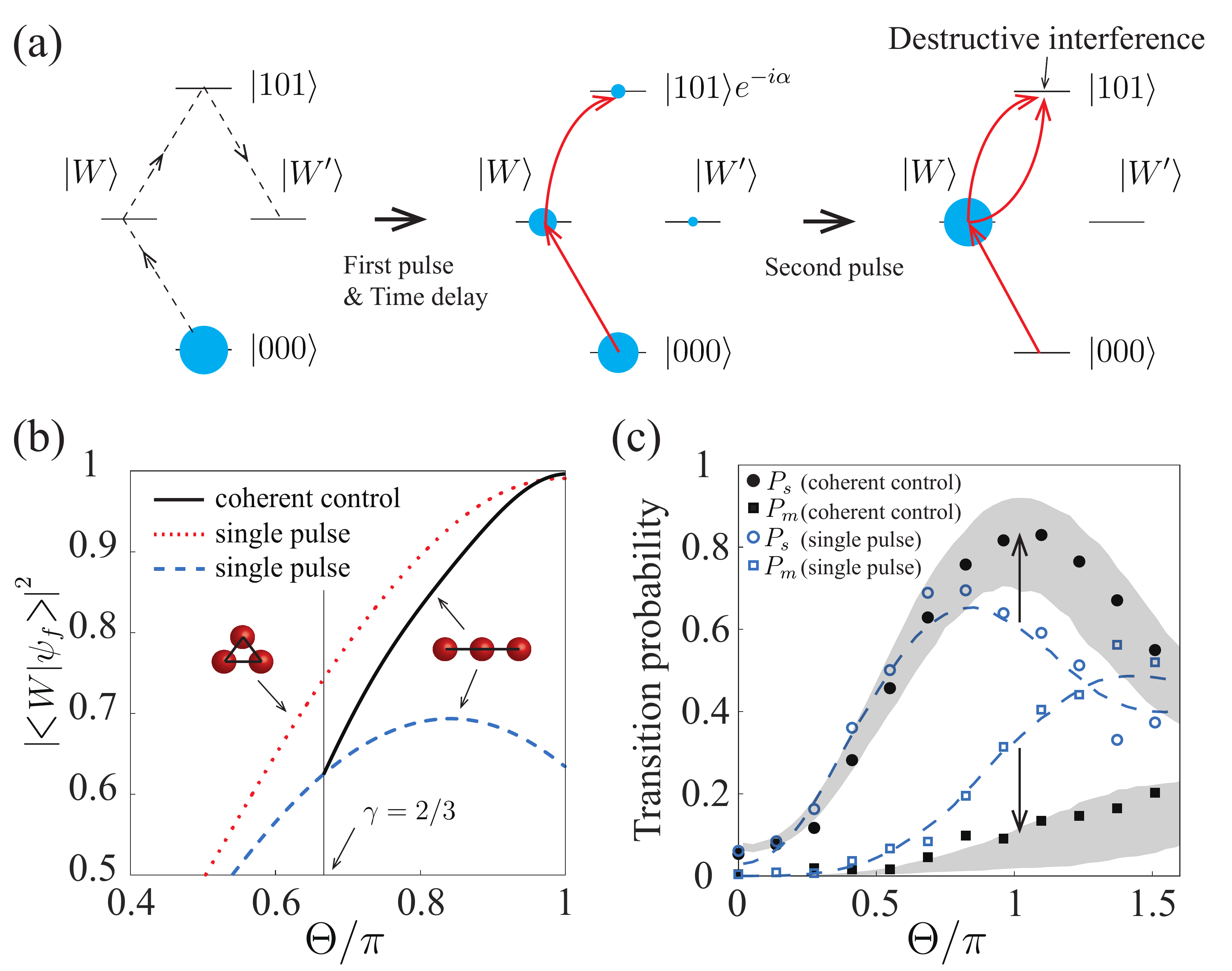}
 \caption{ (a) A coherent control scheme to generate the $\ket{W}$ state of partially-blockaded ($d<r_b<2d$) linear three atoms (see the text for details).  
(b) Transition probability $|\bra{W}\psi_f(\Theta)\rangle|^2$ vs.~pulse area $\Theta$, calculated for the coherent-control case (solid line) with $\ket{\psi_f(\Theta)}=R_y ({\Theta-2\pi/3}) U_p^{AC}(\pi) R_y(2\pi/3) \ket{000}$ for $2\pi/3<\Theta<\pi$. In comparison, single-pulse cases of fully-blockaded triangular three atoms (dotted line) and the given linear three atoms (dashed line) are calculated with $\ket{\psi_f}=R_y (\Theta) \ket{000}$. (c) Measured probabilities $P_{\rm s}$ (of all singly-excited states) and $P_{\rm m}$ (of all multiply-excited states) vs. $\Theta$, in which state preparation and measurement errors about 10\% are taken into account and the shades represent central 50\% sampling ranges of Monte-Carlo simulations.
}
 \label{fig4}
\end{figure}

In the final experiment, we consider an application of the weak coupling for the generation of multi-partite entanglements. For example, a system of partially blockaded three atoms (e.g.,  of $d_{AB},d_{BC}<r_b <d_{AC}$ in the linear configuration), is never driven perfectly to the super-atom state, $\ket{W}=(\ket{100}+\ket{001}+\ket{010})/\sqrt{3}$, by a single resonant excitation of any pulse area $\Theta$, i.e., $R(\Theta)\ket{000} \neq \ket{W}$. Instead, the system also evolves to the leakage states $\ket{W'}=(\ket{100}+\ket{001}-2\ket{010})/\sqrt{6}$ and $\ket{101}$, due to the broken symmetry~\cite{WJL2019}. In order to produce the $\ket{W}$ state of the partially blockaded three atoms, we adopt a coherent control method~\cite{CC} that can destructively interfere unwanted leakage transitions and generate the target state with high fidelity. The first-order leakage state of the given partially-blockade system is $\ket{101}$, so the given leakage transition, $\ket{W}\rightarrow \ket{101}$, can be undone by an additional phase-flipped transition, as illustrated in Fig.~\ref{fig4}(a). Numerical optimization, performed for three atoms with $d_{AB}=d_{BC}=0.66r_b$ and $d_{AC}=1.31r_b$ (in this case $r_b=8.96$~$\mu$m and $\Omega = 2\pi \times 0.99$~MHz), predicts that the given leakage suppression~\cite{HL_leakage} can be engineered, by two pulses of respective pulse-areas $2\pi/3$ and $\pi/3$. All the leakages to other singly-excited and multiply-excited states can be simultaneously suppressed, resulting in high-fidelity $\ket{W}$-state generation of $|\bra{W}R_y ({\pi/3}) U_p^{AC}(\pi) R_y(2\pi/3) \ket{000}|>99.5$\%, as shown in Fig.~\ref{fig4}(b). Experimental results are presented in Fig.~\ref{fig4}(c), in which the probabilities of all singly-excited states, $P_{\rm s}=P_{100}+P_{010}+P_{001}$, and of all multiply-excited states, $P_{\rm m}=P_{110}+P_{011}+P_{101}+P_{111}$, are plotted with filled circles and squares, respectively. Compared with the corresponding single-pulse experiments (open circles and squares), the coherent control scheme shows significant improvements.

Our entanglement scheme (controlled phase gates) can be in principle generalized for $N$ atoms, by utilizing individual atom addressing~\cite{addressing, addressing2}. However, even in this case, maximally available $N$, estimated to be $N_{\rm max} \approx (2d_{\rm max}/d_{\rm min})^3$ in a three-dimensional array, is limited by two distance inequalities: (1) The distance of nearest pairs needs to be larger than the Rydberg blockade radius, i.e., $d_{\rm min} = r_b$, and (2) the dephasing time $T_2$ limits the maximal distance, i.e., $d_{\rm max}=(C_6 T_2/\pi)^{1/6}$. Our current setup with $T_2=5.6$~$\mu$s allows $N_{\rm max}\approx 25$. It is expected that an increased dephasing time (e.g., $4 \times T_2$), by sideband cooling~\cite{sideband_cooling} and a pulse sequence such as an spin echo~\cite{high_fidelity}, and a reduced blockade radius, by an increased Rabi oscillation frequency (e.g., $5 \times \Omega$), shall be able to achieve $N_{\rm max}> 100$.

In conclusion, our observation of Rydberg-atom entanglements in the weak coupling regime demonstrates the remaining Model A of the Rydberg-atom two-qubit gate proposal in Ref.~\cite{Jaksch2000}, which not only completes the proposal, but also allows remote entanglements, larger system applications, and multi-pulse coherent control schemes, as addressed in the paper. The  methods demonstrated in this work may be of particular importance in dealing with entanglements of massive qubit systems, e.g., one-way quantum computing with geometrically imprinted cluster states~\cite{cluster}.

\begin{acknowledgements}
This research was supported by Samsung Science and Technology Foundation [SSTF-BA1301-52] and National Research Foundation of Korea [NRF-2017R1E1A1A01074307]. 
\end{acknowledgements}


\begin{thebibliography}{1}

\bibitem{Horodecki2009} R. Horodecki, P. Horodecki, M. Horodecki, and K. Horodecki, ``Quantum entanglement,'' Rev. Mod. Phys. {\bf 81}, 865 (2009).


\bibitem{Islam2015} R. Islam, R. Ma, P. M. Preiss, M. E. Tai, A. Lukin, M. Rispoli, and M. Greiner, ``Measuring entanglement entropy in a quantum many-body system,'' Nature {\bf 528}, 77 (2015).

\bibitem{Ryd_simulator_lukin} H. Bernien, S. Schwartz, A. Keesling, H. Levine, A. Omran, H. Pichler, S. Choi, A. S. Zibrov, M. Endres, M. Greiner, V. Vuleti\'{c}, and M. D. Lukin, ``Probing many-body dynamics on a 51-atom quantum simulator,'' Nature {\bf 551}, 579–584 (2017).

\bibitem{Ryd_simulator_HSK} H. Kim, Y. J. Park, K. Kim, H.-S. Sim, and J. Ahn, ``Detailed Balance of Thermalization Dynamics in Rydberg-Atom Quantum Simulators,'' Phys. Rev. Lett. {\bf 120}, 180502 (2018).


\bibitem{Nielsen2010} M. A. Nielsen and I. L. Chuang,  \textit{Quantum Computation and Quantum Information}, (Cambridge Univ. Press, 2010).

\bibitem{Degen2017} C. L. Degen, F. Reinhard, and P. Cappellaro, ``Quantum sensing,'' Rev. Mod. Phys. {\bf 89}, 035002 (2017).

\bibitem{Ursin2007} R. Ursin, F. Tiefenbacher, T. Schmitt-manderbach, H. Weier, T. Scheidl, M. Lindenthal, B. Blauensteiner, T. Jennewein, J. Perdigues, P. Trojek, B. Omer, M. Furst, M. Meyenburg, J. Rarity, Z. Sodnik, C. Barbieri, H. Weinfurter, and A. Zeilinger, ``Entanglement-based quantum communication over 144 km,'' Nat. Phys. {\bf 3}, 481 (2007).

\bibitem{entangled_photon} P. G. Kwiat, K. Mattle, H. Weinfurter, A. Zeilinger, A. V. Sergienko, and Y. Shih,  ``New High-Intensity Source of Polarization-Entangled Photon Pairs,'' Phys. Rev. Lett. {\bf 75}, 4337 (1995).

\bibitem{wilk2010} T. Wilk, A. Ga\"{e}tan, C. Evellin, J. Wolters, Y. Miroshnychenko, P. Grangier, and A. Browaeys, ``Entanglement of Two Individual Neutral Atoms Using Rydberg Blockade,'' Phys. Rev. Lett. {\bf 104}, 010502 (2010).

\bibitem{Isenhower2010} L. Isenhower, E. Urban, X. L. Zhang, A. T. Gill, T. Henage, T. A. Johnson, T. G. Walker, and M. Saffman, ``Demonstration of a Neutral Atom Controlled-NOT Quantum Gate,'' Phys. Rev. Lett. {\bf 104}, 010503 (2010).

\bibitem{jau2016} Y.-Y. Jau, A. M. Hankin, T. Keating, I. H. Deutsch, and G. W. Biedermann, ``Entangling atomic spins with a Rydberg-dressed spin-flip blockade,'' Nat. Phys. {\bf 12}, 71 (2016).


\bibitem{high_fidelity} H. Levine, A. Keesling, A. Omran, H. Bernien, S. Schwartz, A. S. Zibrov, M. Endres, M. Greiner, V. Vuleti\'{c}, and M. D. Lukin, ``High-Fidelity Control and Entanglement of Rydberg-Atom Qubits,'' Phys. Rev. Lett. {\bf 121}, 123603 (2018).

\bibitem{entanglement_ion} Q. A. Turchette, C. S. Wood, B. E. King, C. J. Myatt, D. Leibfried, W. M. Itano, C. Monroe, and D. J. Wineland, ``Deterministic Entanglement of Two Trapped Ions,'' Phys. Rev. Lett. {\bf 81}, 3631 (1998).


\bibitem{cnot_supercond} J. H. Plantenberg, P. C. de Groot, C. J. P. M. Harmans, and J. E. Mooij, ``Demonstration of controlled-NOT quantum gates on a pair of superconducting quantum bits,'' Nature {\bf 447}, 836 (2007).

\bibitem{NV_entanglement} P. Neumann, N. Mizuochi, F. Rempp, P. Hemmer, H. Watanabe, S. Yamasaki, V. Jacques, T. Gaebel, F. Jelezko, and J. Wrachtrup, ``Multipartite entanglement among single spins in diamond,'' Science {\bf 320}, 1326 (2008).



\bibitem{Cirac_zoller1} J. I. Cirac and P. Zoller, ``Quantum Computations with Cold Trapped Ions,'' Phys. Rev. Lett. {\bf 74}, 4091 (1995).

\bibitem{Cirac_zoller2} F. Schmidt-Kaler, H. H\"{a}ffner, M. Riebe, S. Gulde, G. P. T. Lancaster, T. Deuschle, C. Becher, C. F. Roos, J. Eschner, and R. Blatt, ``Realization of the Cirac-Zoller controlled-NOT quantum gate,'' Nature {\bf 422}, 408–411 (2003).

\bibitem{entalgment_supercond} A. J. Berkley, H. Xu, R. C. Ramos, M. A. Gubrud, F. W. Strauch, P. R. Johnson, J. R. Anderson, A. J. Dragt, C. J. Lobb, and F. C. Wellstood, ``Entangled macroscopic quantum states in two superconducting qubits,'' Science {\bf 300}, 1548 (2003).

\bibitem{Jaksch2000} D. Jaksch, J. I. Cirac, P. Zoller, S. L. Rolston, R. Cote, and M. D. Lukin, ``Fast Quantum Gates for Neutral Atoms,'' Phys. Rev. Lett. {\bf 85}, 2208 (2000).



\bibitem{entanglement_measurement2} A. Widera, O. Mandel, M. Greiner, S. Kreim, T. W. H\"{a}nsch, and I. Bloch, ``Entanglement Interferometry for Precision Measurement of Atomic Scattering Properties,'' Phys. Rev. Lett. {\bf 92}, 160406 (2004).

\bibitem{tweezer} H. Kim, W. Lee, H. Lee, H. Jo, Y. Song, and J. Ahn, ``In situ single-atom array synthesis using dynamic holographic optical tweezers,'' Nat. Comm. {\bf 7}, 13317 (2016).

\bibitem{WJL2019} W. Lee, M. Kim, H. Jo, Y. Song, and J. Ahn, ``Coherent and dissipative dynamics of entangled few-body systems of Rydberg atoms,'' Phys. Rev. A {\bf 99}, 043404 (2019).

\bibitem{pairint} S. Weber, C. Tresp, H. Menke, A. Urvoy, O. Firstenberg, H. P. B\"{u}chler, and S. Hofferberth, ``Tutorial: Calculation of Rydberg interaction potentials,'' J. Phys. B: At. Mol. Opt. Phys. {\bf 50}, 133001 (2017).

\bibitem{CC} C. Brif, R. Chakrabarti, and H. Rabitz,
``Control of quantum phenomena: past, present and future,''
New J. Phys. {\bf 12},  075008 (2010).

\bibitem{HL_leakage} H. Jo, Y. Song, and J. Ahn, ``Qubit leakage suppression by ultrafast composite pulses,'' Opt. Express {\bf 27}, 3944 (2019).


\bibitem{addressing} H. Labuhn, S. Ravets, D. Barredo, L. B\'{e}guin, F. Nogrette, T. Lahaye, and A. Browaeys, ``Single-atom addressing in microtraps for quantum-state engineering using Rydberg atoms,'' Phys. Rev. A {\bf 90}, 023415 (2014).

\bibitem{addressing2} Y. Wang, X. Zhang, T. A. Corcovilos, A. Kumar, and D. S. Weiss, ``Coherent Addressing of Individual Neutral Atoms in a 3D Optical Lattice,'' Phys. Rev. Lett. {\bf 115}, 043003 (2015).

\bibitem{sideband_cooling} A. M. Kaufman, B. J. Lester, and C. A. Regal, ``Cooling a single atom in an optical tweezer to its quantum ground state,'' Phys. Rev. X {\bf 2}, 041014 (2012).


\bibitem{cluster} R. Raussendorf and H. J. Briegel, ``A One-Way Quantum Computer,'' Phys. Rev. Lett. {\bf 86}, 5188 (2001).



\end{thebibliography}
\end{document}